\documentclass[conference]{IEEEtran}
\usepackage{hyperref}
\IEEEoverridecommandlockouts
\usepackage{cite}
\usepackage{amsmath,amssymb,amsfonts}
\usepackage{algorithmic}
\usepackage{graphicx}
\usepackage{textcomp}
\usepackage{xcolor}
\usepackage{stfloats}
\graphicspath{{../figs/}}
\usepackage{caption}
\usepackage{listings}
\usepackage{subcaption}
\DeclareGraphicsExtensions{.png}
\usepackage{float}
\def\BibTeX{{\rm B\kern-.05em{\sc i\kern-.025em b}\kern-.08em
    T\kern-.1667em\lower.7ex\hbox{E}\kern-.125emX}}

\begin{document}

\newcommand{\leo}[1]{\textcolor{red}{#1}}
\newcommand{\luca}[1]{\textcolor{red}{#1}}

\title{Resource Analysis of Ethereum 2.0 Clients}

%\thanks{Research financed by the Ethereum foundation}

\author{\IEEEauthorblockN{Mikel Cortes-Goicoechea}
\IEEEauthorblockA{\textit{Barcelona Supercomputing Center} \\
Barcelona, Spain \\
mikel.cortes@bsc.es}
\and
\IEEEauthorblockN{Luca Franceschini}
\IEEEauthorblockA{\textit{Barcelona Supercomputing Center} \\
Barcelona, Spain \\
luca.franceschini@bsc.es}
\and
\IEEEauthorblockN{Leonardo Bautista-Gomez}
\IEEEauthorblockA{\textit{Barcelona Supercomputing Center} \\
Barcelona, Spain \\
leonardo.bautista@bsc.es}

}

\maketitle
\thispagestyle{plain}
\pagestyle{plain}

\begin{abstract}
Scalability is a common issue among the most used permissionless blockchains, and several approaches have been proposed accordingly. As Ethereum is set to be a solid foundation for a decentralized Internet web, the need for tackling scalability issues while preserving the security of the network is an important challenge. In order to successfully deliver effective scaling solutions, Ethereum is on the path of a major protocol improvement called Ethereum 2.0 (Eth2), that implements sharding.  As the change of consensus mechanism is an extremely delicate matter, this improvement will be achieved through different phases, first of which is the implementation of the Beacon Chain. For this, a specification has been developed and multiple groups have implemented clients to run the new protocol. In this work, we analyse the resource usage behaviour of different clients running as Eth2 nodes, comparing their performance and analysing differences. Our results show multiple network perturbations and how different clients react to it.
\end{abstract}


\section{Introduction}

Ethereum~\cite{eth-whitepaper} has been a great achievement in the road to ubiquitous blockchain technology. It led to a huge growth in the number of decentralized applications, due to its general purpose virtual machine and its dedicated programming language. These characteristics have set the conditions for a solid community of developers and continuous advancements as well as introducing new technological possibilities. As the Ethereum adoption increases, its usability has been threatened by the rising transactions volume and network clogging. 
%Due to the fact that block space is a scarce resource, the constant rise in network usage provokes transaction fees to increase considerably, preventing users from interacting with the network for a reasonable price. 

In order to successfully implement effective scaling solutions, Ethereum is on the path to a major protocol improvement, that will enhance its scalability by several orders of magnitude and provide an architecture to flexibly address the needs of a constantly changing industry. Ethereum 2.0 (Eth2) is based around the concept of sharding, where the blockchain is split into shards and subsets of validators are randomly assigned to each shard in order to validate transactions. As validators just need to validate transactions relative to the shards they have been assigned to, parallelism and therefore scalability is achieved. To have a coherent state of the network between different shards, the roots of the shard blocks are appended to a dedicated chain called the Beacon Chain. This Beacon Chain is considered to be the \emph{heart beat} and the foundation of the sharding protocol. Validators shuffle randomly every epoch, verifying transactions on different shards. 

In order to embrace these changes to the final stage of Eth2, the consensus mechanism of Ethereum will change from Proof-of-Work (PoW) to Proof-of-Stake (PoS). This change allows blocks to be produced in a more sustainable way, saving electricity while implementing a more exhaustive network infrastructure. Due to the complexity of introducing this consensus mechanism, the achievement of sharding in Eth2 has been split into different phases. In the first phase, called “Phase 0”\cite{phase-0}, PoS is implemented on the Beacon Chain and validators participate as part of proposers and attestation committees. The Beacon Chain provides the random assignment of validators to committees, and this sets the foundation of the next phases. 

%The Eth2 Phase 0 specification has been developed and a number of groups have launch themselves in the quest of developing an Eth2 client fully compliant with the Phase 0 specification. As of today (November 2020), five clients have achieved this mission (or are close to) and anyone can download and run their clients in order to participate in the Eth2 network. In this paper we analyze the resource utilization and performances of the five different Eth2 clients, participating as nodes of the Medalla Testnet.

The remainder of this paper is organized as follows. 
Section~\ref{sec:methodology} explains the methodology used during for evaluation.
In Section~\ref{sec:analysis} we show and analyse the results obtained by our study.
Section~\ref{sec:relatedwork} discusses related work.
Finally, Section~\ref{sec:conclusion} concludes this work and presents some possible future directions.
\section{Methodology}
\label{sec:methodology}

In order to study differences in behaviour of Eth2 clients, we have been running the Eth2 clients Teku, Nimbus, Lighthouse, Prysm and Lodestar for several hours, letting the clients sync to the Eth2 chain from scratch. The objective of this study is to monitor specific metrics in order to understand the behaviour and performance of the clients when initialized to sync to the Eth2 network. Clients have not always been run at the same time, as Prysm and Nimbus were launched and ran together, while Teku, Lighthouse and Lodestar have been run during different periods of time. In addition to some differences in clients’ behaviour that may have been caused by conditions of the Medalla Test Network\cite{medallanet} at a given time, we have found out some patterns and gathered some insights about the clients. The metrics that have been monitored are:
\begin{itemize}
\item Syncing time
\item Peer connections
\item Network outgoing traffic
\item Network incoming traffic
\item Memory Usage
\item CPU
\item Disk Usage
\end{itemize}

The metrics mentioned above have been collected by launching the clients with an empty database and no external processes running. After having located the client process ID, we launched a resource monitoring script written in python~\cite{metrics-monitoring-scripts} that records the resource utilization every second into a data file. While launching the clients, a flag has been added in order to save clients' logs to a text file. The logs were parsed through python scripts~\cite{logs-parsing-scripts}, allowing to extrapolate the syncing speed of the clients. 
Metrics and slot times were plotted through python-matplotlib scripts~\cite{plotting-scripts}, and saved into CSV files.

\subsection{Evaluation platform}
\label{sec:platform}

Tests have been run on two nodes, each node with an Intel(R) Xeon(R) CPU E5-2620 0 @ 2.00GHz with 1 core and two threads.
Each client has been run on a single node with 4.8 GB of total memory available and 34GB of storage (40GB total, of which 6GB used for the operating system).

Clients have been run on the Medalla Test Network using their default configuration. No additional flags or configuration have been used with the exception of the flag on Teku that limits the JVM memory and rising the heap size of Lodestar on the package.json. Teku ran allocating 2 GB to the JVM, as allocating more than 2GB would make the server crash. Similar issues occurred with Lodestar. 

\subsection{Client versions and running period}
\label{sec:clients}

Clients have been run with the following versions:
\begin{itemize}
\item Teku: v0.12.14-dev-6883451c \cite{teku-version}
\item Prysm: v1.0.0-beta.1-4bc7cb6959a1ea5b\cite{prysm-version}
\item Lighthouse: v0.3.0-95c96ac5\cite{lighthouse-version}
\item Nimbus: 0.5.0-9255945f\cite{nimbus-version}
\item Lodestar: commit 40a561483119c14751 \cite{lodestar-version}
\end{itemize} 

%\subsection{Client running period}

Clients have been running during the periods depicted in Table~\ref{tab:runtime}. Teku and Lighthouse were run several times in order to compare the resource usage of two different executions.

\begin{table}[ht]
 \caption{Client Running Periods}
 \label{tab:runtime}
\begin{center}
\begin{tabular}{ |c|c|c| } 
 \hline
 Client & Start Time & End Time \\ 
 \hline
 Teku & 2020-11-09 17:25:12 & 2020-11-10 17:34:45 \\ 
 Prysm & 2020-11-04 18:34:12 & 2020-11-06 09:34:34 \\ 
 Lighthouse & 2020-11-02 17:17:51 & 2020-11-04 02:57:38 \\
 Nimbus & 2020-11-04 18:40:35 & 2020-11-06 10:23:04 \\
 Lodestar & 2020-11-08 20:19:02 & 2020-11-09 08:54:04 \\
 \hline
\end{tabular}
\end{center}
\end{table}

\section{Analysis}
\label{sec:analysis}

In this section we go over the results obtained during the runs of the five clients and we analyse their differences in resource consumption and behaviour.

\subsection{Client syncing}

First, we start by studying the synchronization time of all clients. It is important to note that the synchronization time of a client is not really a determining factor of its efficiency. First, this is a cost that it should be paid only once in the lifetime of a client (assuming no hard failures that force a restart from scratch), but also because most clients can now start from a weak subjectivity state and be up and running in under a minute. Nonetheless, synchronization time is very relevant when comparing it with other resources metrics to try to uncover hidden effects or understand unexpected behaviour and this is the way we use it in this research. Other studies have also used syncing metrics in a similar way~\cite{afri}.

\begin{figure}[H]
\centerline{\includegraphics[scale=0.36]{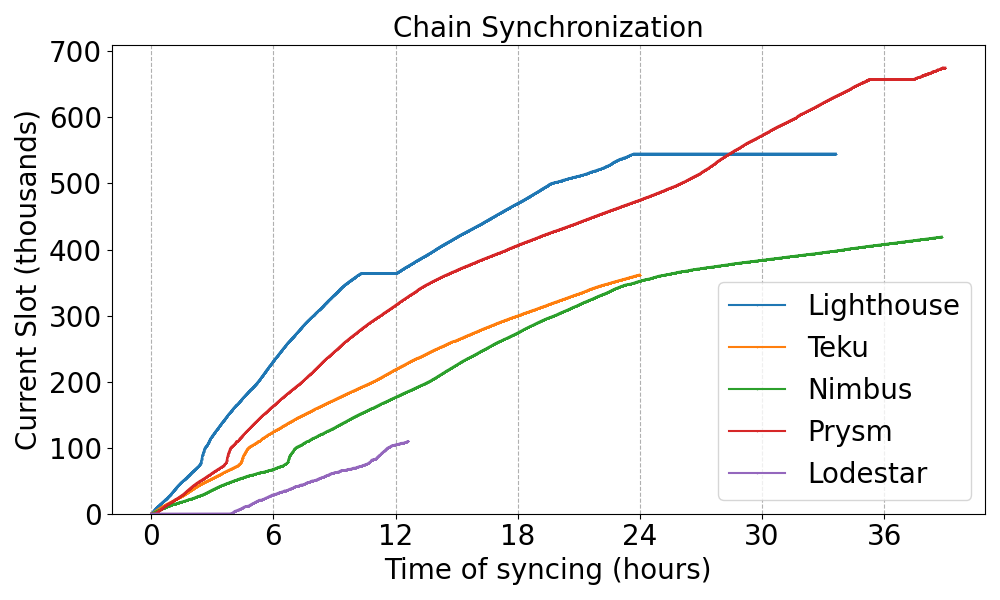}}
\caption{Slot synchronization of Eth2 clients}
\label{fig:clients-current-slot}
\end{figure}

Figure \ref{fig:clients-current-slot} shows the synchronization slot as the client runs from genesis. We tried to run all clients for over 24 hours, with the exception of Lodestar, which suffered from multiple crashes and had to be restarted several times. We can see that Lighthouse appears to be the fastest syncing client, though its syncing process stops around hour 24 due to reaching the maximum storage capacity of the server (See Section~\ref{sec:disk}). After Lighthouse stopped its syncing process, Prysm was able to catch up and sync faster than the other clients with a steady progress throughout the whole test. Nimbus and Lodestar seem to be the ones that take more time to sync. More importantly, it can be seen that all clients experience a sharp increase in syncing speed around slot 100,000. The fact that all clients observe the same change of steepness in the syncing curve around this slot might suggest that this phenomena is inherited from the network conditions (See Section ~\ref{sec:fossil}).

\subsection{Peer connections}
\label{sec:peers}

\begin{figure}[H]
\centerline{\includegraphics[scale=0.36]{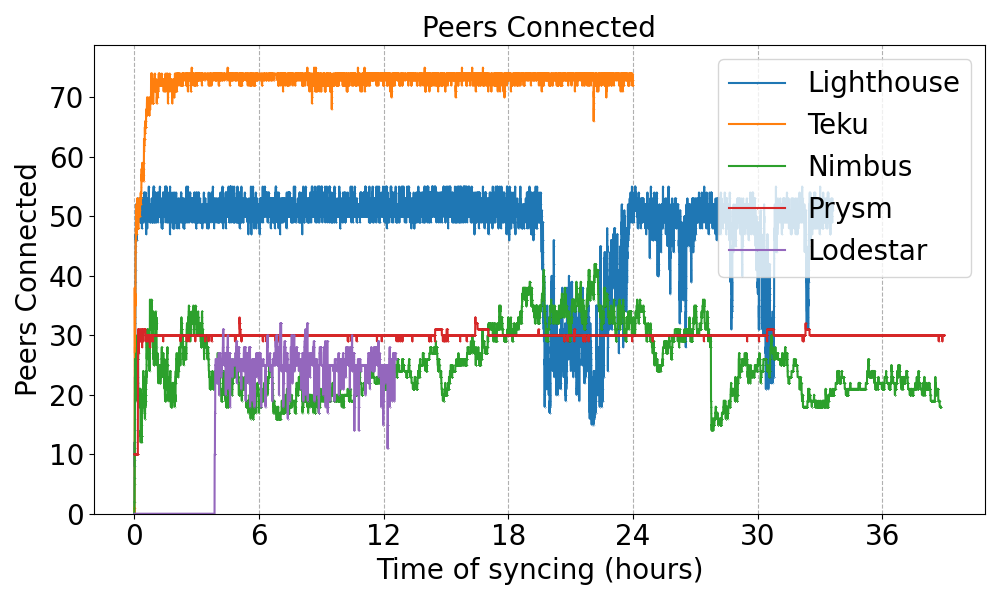}}
\caption{Peer connections of Eth2 Clients}
\label{fig:clients-peers-connected}
\end{figure}

We kept track of the number of peer connections for each client during the experiment. While this is a parameter that can be tuned, we used the default values for each client, as this is what most user would do on a regular basis. We can see in Figure~\ref{fig:clients-peers-connected} that Teku connects to the highest number of peers (i.e., over 70 peers) than any other client and keeps a stable number of connections during the entire experiment. Lighthouse is the client that connects with more peers after Teku with about 50 peers. After 20 hours of execution, the number of connected peers drastically decreased (oscillating around 25 peers), and later rises its peer connection back to 50 peers (hour 24). After hour 24 Lighthouse experienced sharp drops and recoveries in peer connections. It is important to notice that Lighthouse was the first client to be run, from November 2dn to November 4th, while the Medalla Testnet was experiencing some issues related to non-finality~\cite{medalla-nonfinality-october} which might explain the peer connection drops as well as other strange phenomena observed in the Lighthouse run. This allowed us to do a more detailed analysis of the data. Nimbus peers connections oscillate significantly, usually connecting with over 15 peers and arriving up to 42 peers. Prysm distinguishes itself for its stability of peer connections, as it succeeds in achieving 30 peer connections with almost no variation. When it comes to Lodestar, the client does not report the number of peers until the moment in which it retrieves the genesis block (hour 4), and from this point Lodestar’s peer connections oscillate in the range of 20-30 peers.

\subsection{Outgoing network traffic}
\label{sec:netout}

\begin{figure}[h]
\centerline{\includegraphics[scale=0.36]{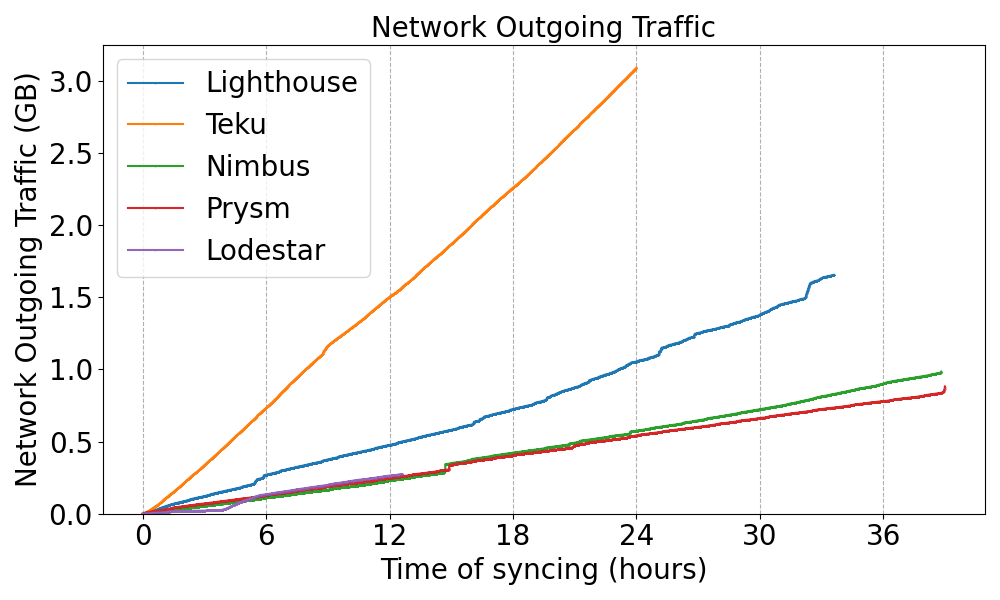}}
\caption{Clients outgoing network traffic}
\label{fig:clients-network-outgoing}
\end{figure}

\begin{figure}[h]
\centerline{\includegraphics[scale=0.36]{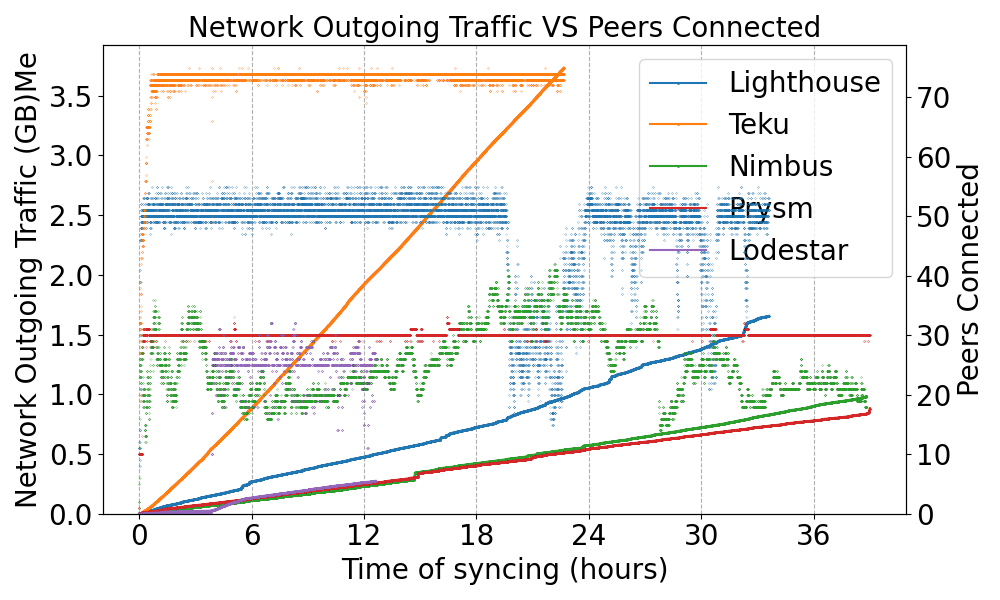}}
\caption{Clients outgoing network traffic and connected peers}
\label{fig:clients-network-outgoing-peers}
\end{figure}

We also monitored the outgoing network traffic for all clients. In general, we can see in Figure \ref{fig:clients-network-outgoing} a steady constant behaviour across all the clients for network outgoing data. Prysm and Nimbus are the most conservative in outgoing network traffic and their behaviour is quite similar (as their network outgoing traffic almost overlaps). Teku seems to share more data than the other clients. In general, the amount of data shared by clients seems to correlate well to the number of peer connections: clients with more peer connections have more outgoing traffic as shown in Figure \ref{fig:clients-network-outgoing-peers}.

\subsection{Incoming Network Traffic}
\label{sec:netin}

\begin{figure}[h]
\centerline{\includegraphics[scale=0.36]{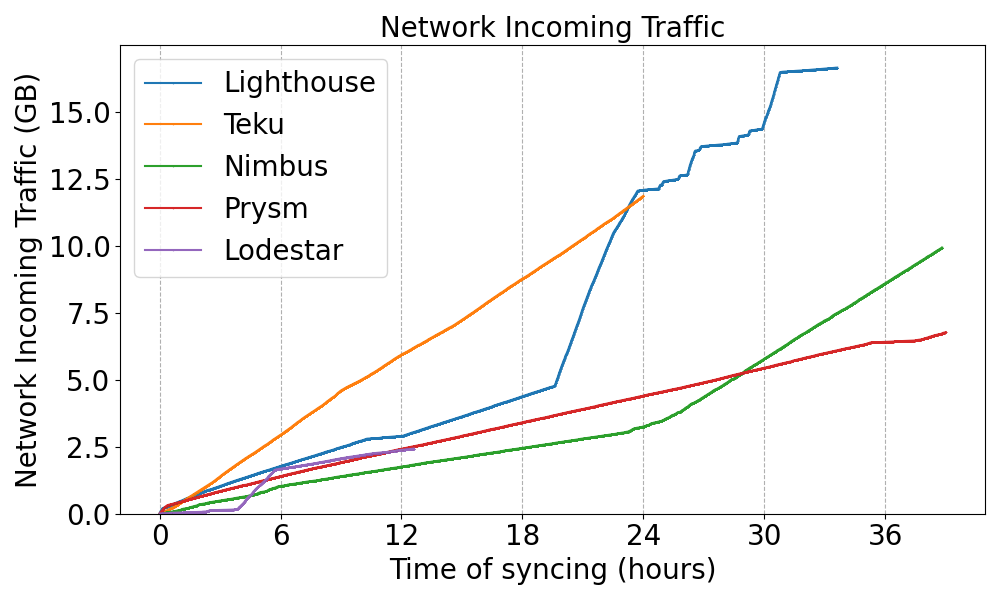}}
\caption{Clients incoming network traffic}
\label{fig:clients-network-incoming}
\end{figure}

For incoming network traffic, we see a much less stable behaviour than for outgoing. We can see in Figure~\ref{fig:clients-network-incoming} that Nimbus and Prysm are again the lightest clients in terms of incoming network traffic. Prysm and Teku show a normal steady behaviour but Nimbus and Lighthouse do not.

\begin{figure}[h]
\centerline{\includegraphics[scale=0.35]{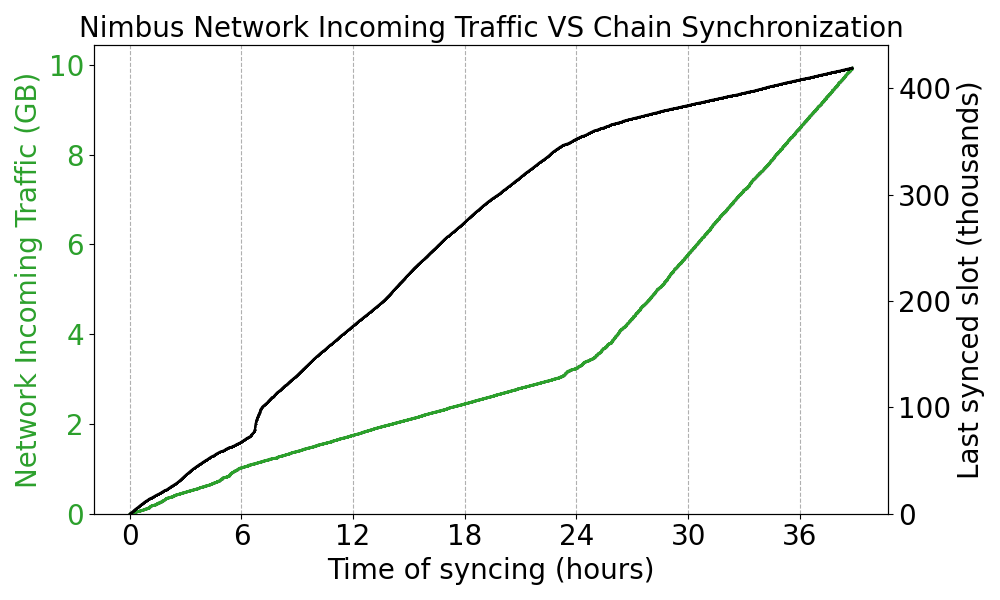}}
\caption{Nimbus incoming traffic (green) and syncing (black)}
\label{fig:nimbus-incoming-traffic}
\end{figure}

\begin{figure}[h]
\centerline{\includegraphics[scale=0.35]{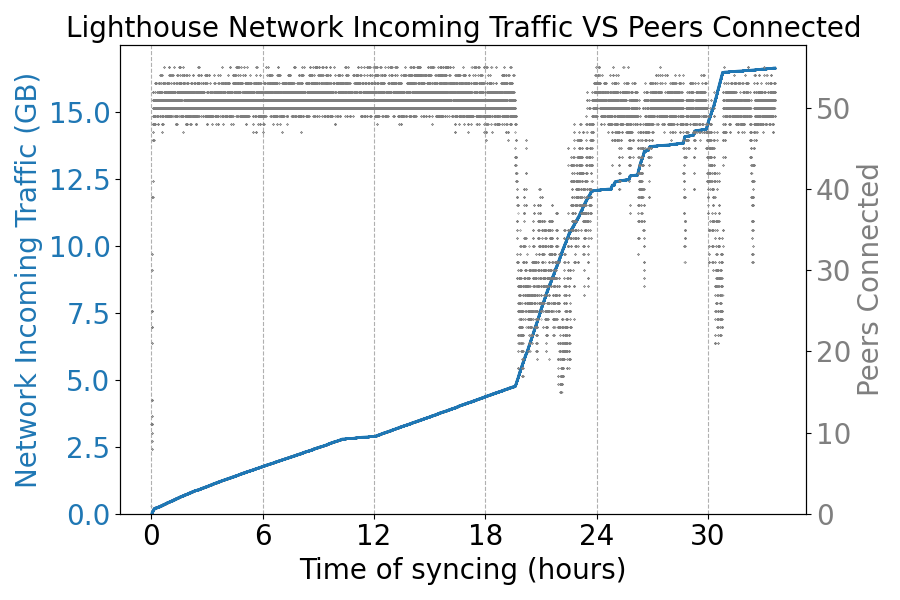}}
\caption{Lighthouse incoming traffic (blue) and peer connections (black)}
\label{fig:lighthouse-network-incoming-peers}
\end{figure}

Nimbus presents an interesting behaviour. As shown in Figure~\ref{fig:nimbus-incoming-traffic}, there is a point in which incoming traffic increases substantially, while at the same time the curve that represents its slot synchronization does not appear to have accelerated accordingly but on the contrary, it slows down. Teku is the client that has more network traffic for almost all of its execution, but Lighthouse sharply increases network incoming data from hour 20 to hour 24, overtaking Teku. To analyse this phenomena in more detail we plot the net incoming data vs the number of peer connections in Figure~\ref{fig:lighthouse-network-incoming-peers}. It can be seen a considerable difference in Lighthouse's incoming network traffic, as we can witness the acceleration in receiving data from 250MB/h (hours 15-17) to 1700MB/h (hours 21-23). We notice that during those hours (i.e., 20-24) the number of peer connections decreases by about 50\%, while the network incoming data increases by a factor of $6.8$. A similar (but smaller) effect is observed in hours 30-32, followed by a period of stability.

Such a behaviour could be reasonably explained by the non-finality period experienced at the end of October and beginning of November, which fits well with the important number of clients that drop out of the network during those hours. This demonstrate, how non-finality periods can impact the clients.

\subsection{Memory}
\label{sec:memory}

\begin{figure}[h]
\centerline{\includegraphics[scale=0.36]{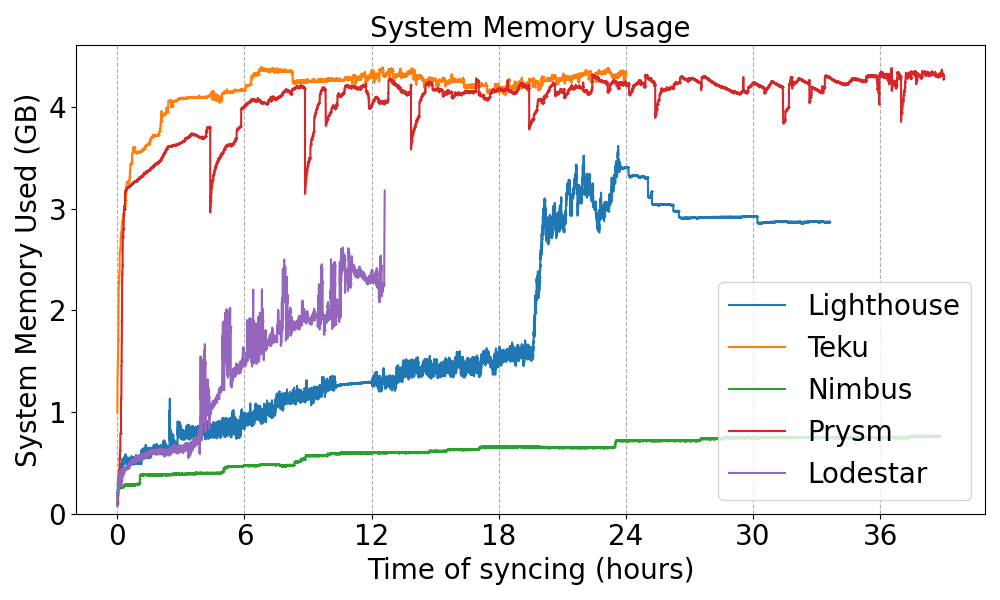}}
\caption{Memory consumption of Eth2 clients}
\label{fig:clients-memory}
\end{figure}

When it comes to memory consumption, we can see in Figure \ref{fig:clients-memory} how Teku and Prysm tend to have a similar behaviour, by starting with a steep usage of memory consumption and settling with a similar high memory consumption in the long run. Interestingly, we can notice sharp drops of memory usage for Prysm at somehow regular intervals, indicating some kind of periodic cleaning process, while the behaviour of Teku is more constant. Nimbus distinguishes itself with a very low memory consumption with minimal changes. Lodestar memory usage is characterized by periodic spikes, and appears to be less demanding than the memory usage of Teku and Prysm. It can also be noticed that the first relevant spike corresponds to the moment in which the client actually starts the syncing process. Lighthouse memory consumption is low overall, but it showed some unstable behaviour.

\begin{figure}[h]
\centerline{\includegraphics[scale=0.38]{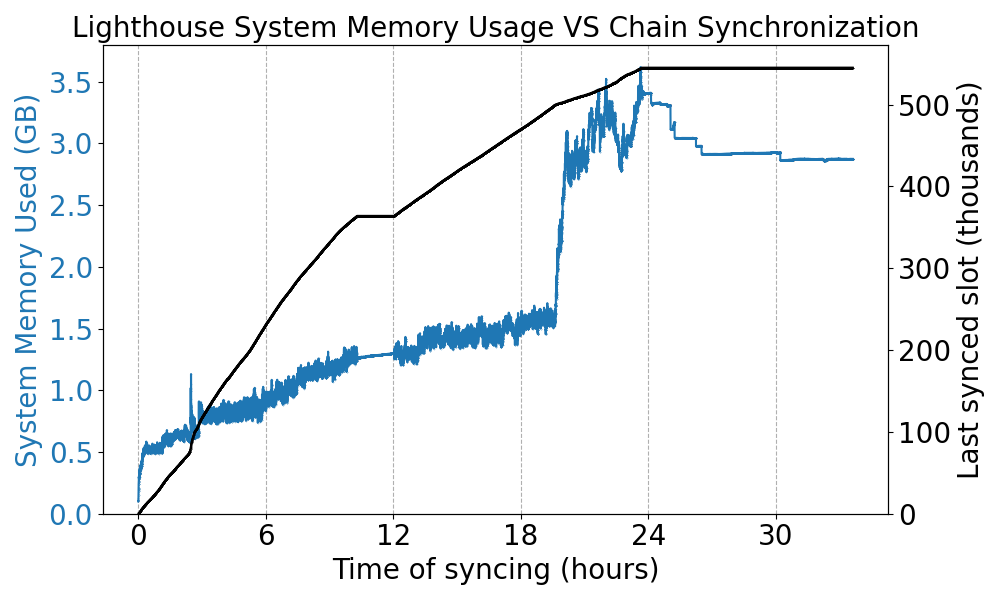}}
\caption{ Lighthouse memory (blue) and synchronization (black)}
\label{fig:lighthouse-memory-slot}
\end{figure}

In Figure \ref{fig:lighthouse-memory-slot} we can witness a considerable rise in memory usage for Lighthouse (hour 20) that corresponds to the time in which Lighthouse began to sharply increase its incoming network traffic and the decrease in peer connections. The memory usage oscillates around 3GB for several hours, until the 24th hour where it can be seen that Lighthouse’s memory usage begins to diminish and then becomes flat. This is due to the storage limitations in the node.
%This last segment corresponds to the point where Lighthouse stops syncing}.

\subsection{CPU}
\label{sec:cpu}

\begin{figure}[h]
\centerline{\includegraphics[scale=0.36]{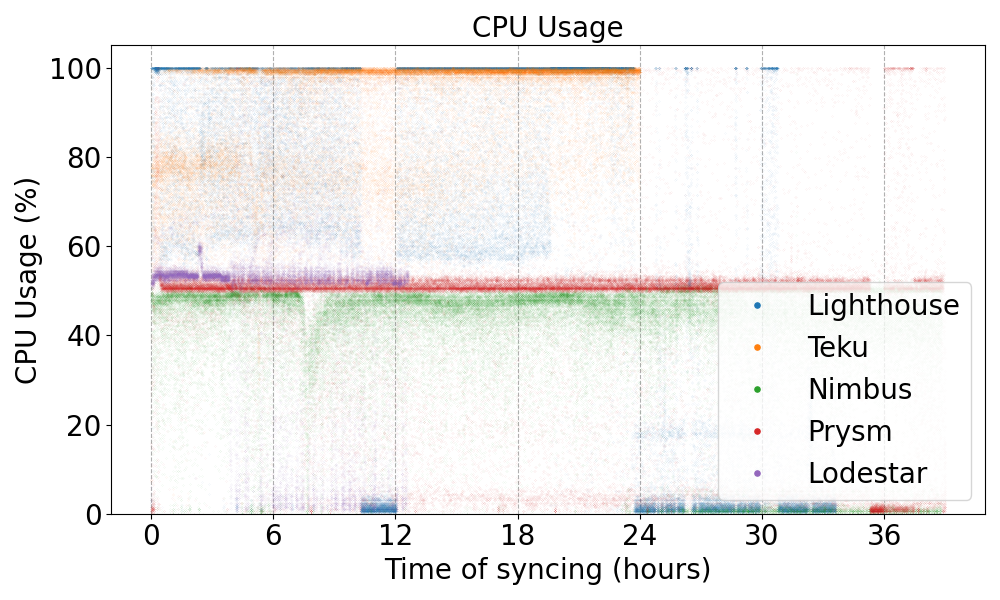}}
\caption{CPU usage of Eth2 clients}
\label{fig:client-cpu}
\end{figure}

From monitoring the CPU usage across the different clients, we noticed in Figure~\ref{fig:client-cpu} that Nimbus has the lowest CPU consumption overall, staying always under 50\%. The CPU usage of Prysm and Lodestar concentrates around 50\%, while the CPU usage of Lighthouse is most of the time above 50\% and Teku constantly oscillates near 100\%. Teku's CPU usage is always high and stable in our experiments. There might be several reasons for this. First, we performed our experiments on a virtual machine with just one core of an Intel(R) Xeon(R) E5-2620 CPU (See Section~\ref{sec:methodology}), which might be significantly more limited than a contemporary node with a more recent CPU generation and multiple cores. Second, we limited the JVM heap size to only 2GB because larger JVM heap sizes kept leading to crashes. With 2GB for the JVM heap, Teku managed to run smoothly without issues. However, 2GB is quite tight for Medalla testnet and it is possible that a significant portion of the time was spent in garbage collection, leading to the observed high CPU utilization.

\begin{figure}[h]
\centerline{\includegraphics[scale=0.38]{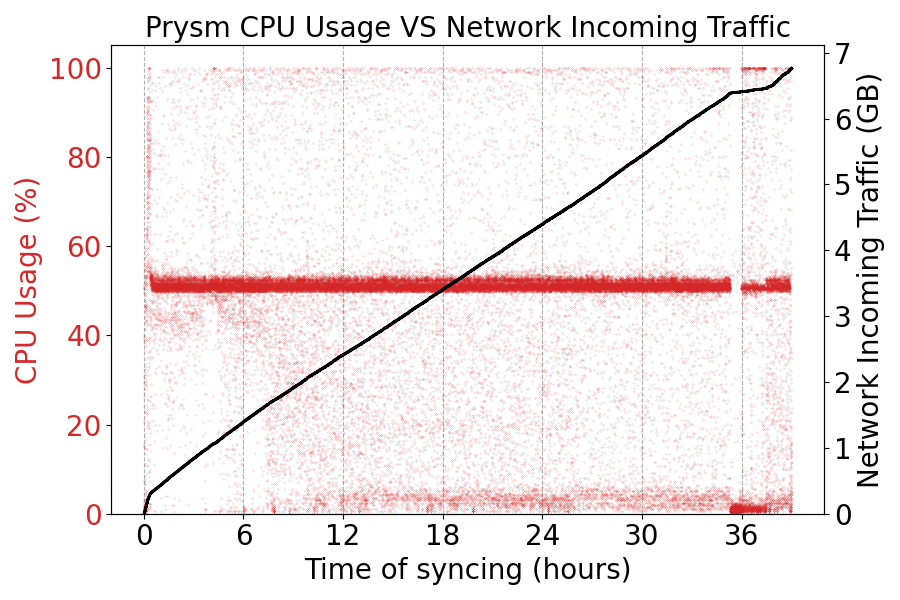}}
\caption{ Prysm CPU (red) and incoming traffic (black)}
\label{fig:prysm-cpu-netin}
\end{figure}

Prysm also shows a relatively constant CPU utilization except for hour 35, where it drops to 0, as shown in Figure \ref{fig:prysm-cpu-netin}. During that time, we also observe a strong reduction on network incoming traffic, which explains the drop on CPU utilization. It is interesting to notice that this happens even while Prysm keeps stable peer connections (See Figure~\ref{fig:clients-peers-connected}).

%\begin{figure}[h]
%\centerline{\includegraphics[scale=0.38]{figs/prysmCPU-DISK.png}}
%\caption{Prysm CPU (red) and disk storage (black)}
%\label{fig:prysm-cpu-disk}
%\end{figure}

\begin{figure}[h]
\centerline{\includegraphics[scale=0.36]{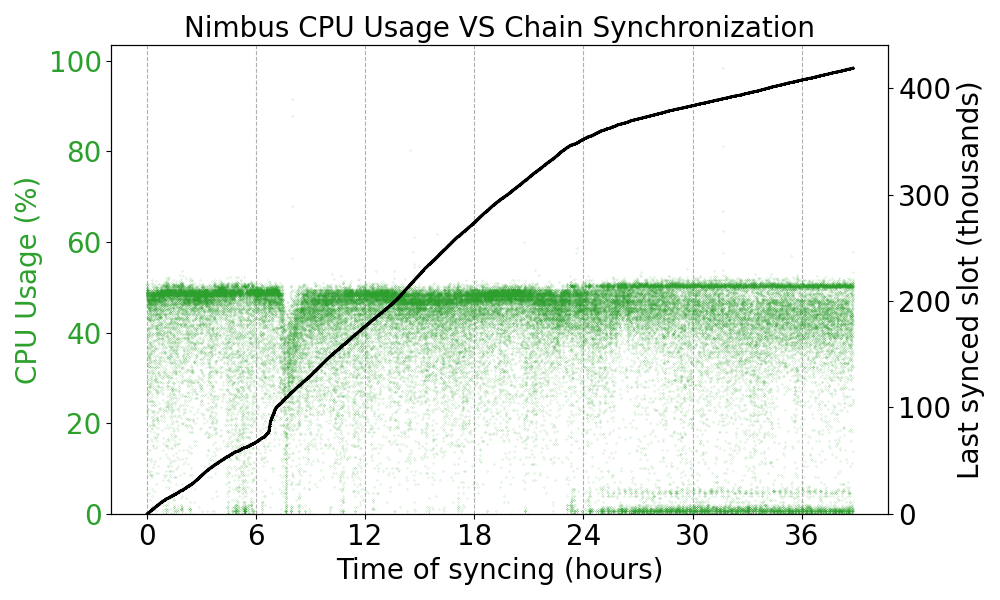}}
\caption{Nimbus CPU (green) and slot synchronization (black)}
\label{fig:nimbus-cpu-slot}
\end{figure}

Nimbus also showed a constant CPU utilization most of the time, except for a short period of time around hour 7 where it witnessed a quite sharp drop in CPU usage, as it can be seen in Figure~\ref{fig:nimbus-cpu-slot}. We also can observe that this drop comes just after a steep increase in blocks synced. In the other clients, such an increase in synced blocks in a period of time is not followed by a CPU drop. This drop in CPU usage will be further investigated later (See Section~\ref{sec:disk}).

\begin{figure}[h]
\centerline{\includegraphics[scale=0.38]{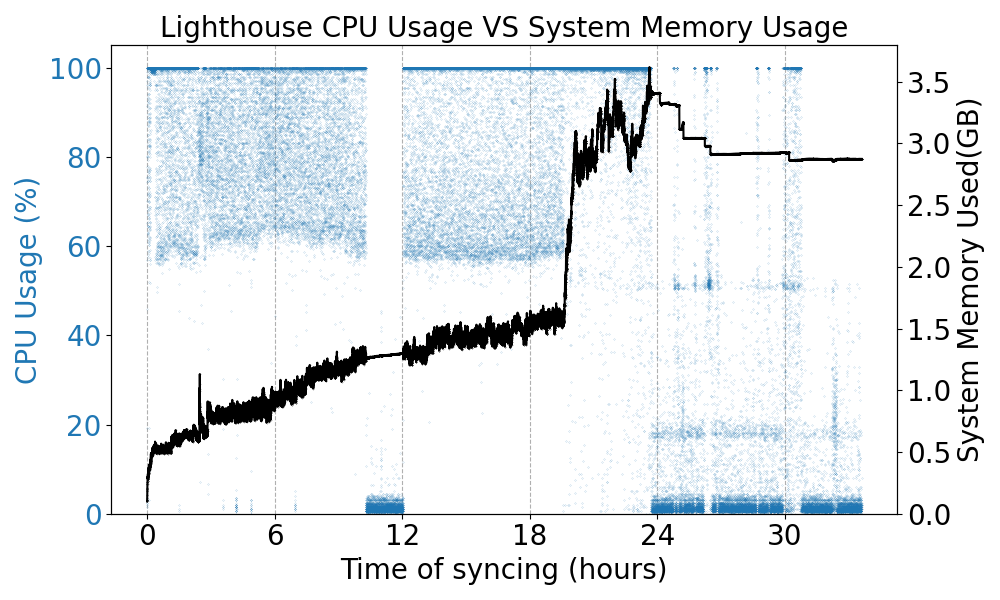}}
\caption{Lighthouse CPU (blue) and memory usage (black)}
\label{fig:lighthouse-cpu-mem}
\end{figure}

Lighthouse showed some irregular CPU utilization pattern with a sudden drop to 0\% CPU utilization accompanied with a flat memory usage during hours 10-12 as shown in Figure~\ref{fig:lighthouse-cpu-mem}. The same happens after hour 24, where the client stops syncing and the CPU utilization spend most of the time near 0\%. After a sharp increase in memory usage during the previous period of dropping peer connection (See Section~\ref{sec:memory}), the memory starts to decrease gradually, ending on a flat line.

\subsection{Disk Usage}
\label{sec:disk}

\begin{figure}[h]
\centerline{\includegraphics[scale=0.34]{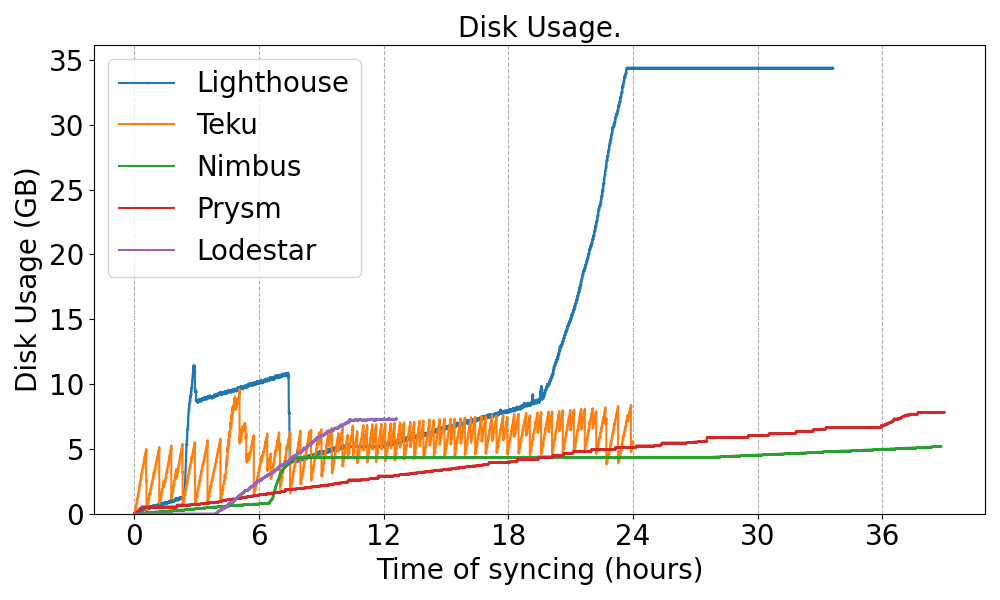}}
\caption{ Disk Usage of Eth2 clients}
\label{fig:clients-disk}
\end{figure}

\begin{figure}[h]
\centerline{\includegraphics[scale=0.35]{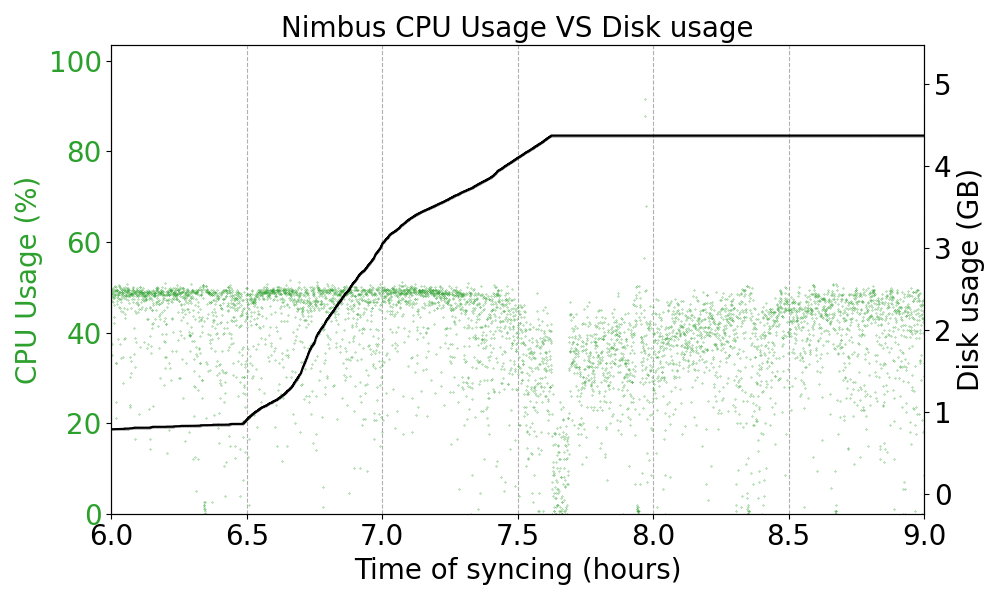}}
\caption{Nimbus CPU (green) and slot synchronization (black)}
\label{fig:nimbus-cpu-disk}
\end{figure}

We plot the disk usage of all clients in Figure~\ref{fig:clients-disk}. This plot shows some unexpected behaviour for some of the clients. The most stable behaviour, in terms of disk usage among all clients, seems to be Prysm according to this experiment. Nimbus maintains a low storage usage for most of the time, although it does experience a sharp increase in storage at some point. We investigate this increase in storage by isolating Nimbus' disk utilization and comparing it with its CPU usage. We can see in Figure~\ref{fig:nimbus-cpu-disk} that the increase in storage is quite dramatic, it goes from 1GB to over 4GB in a short period of time. %LEO: How short?
Curiously, just after this storage increase, the CPU usage of Nimbus drops substantially. While it is true that periods of high storage intensity are often followed by idle CPU time, note that the x-axis is denominated in hours. Such idle CPU times just last for milliseconds or seconds after storage bursts, so this solely it is unlikely to explain the whole drop of CPU utilization. In addition, this is not observed in other clients that also experience large increases in disk utilization.

Lighthouse also shows a steep increase in disk usage, although much more important than Nimbus, going from 3GB to over 10GB in minutes. After this, the disk usage continues to increment at the same speed than initially doing, keeping the same growth rate for several hours, to then drop sharply at around hour 8. Interestingly, extrapolating the initial storage growth rate would place Lighthouse at the same spot after 9 hours of syncing, as if the whole storage used during the perturbation was just temporary data that once everything solved, could be discarded without losing any real information. After 20 hours, however, Lighthouse observes another sharp increase in disk usage, but this time it keeps going up until it consumes the whole 34GB of space in the node, at which point it becomes a flat line for the next 10 hours.

\begin{figure}[h]
\centerline{\includegraphics[scale=0.38]{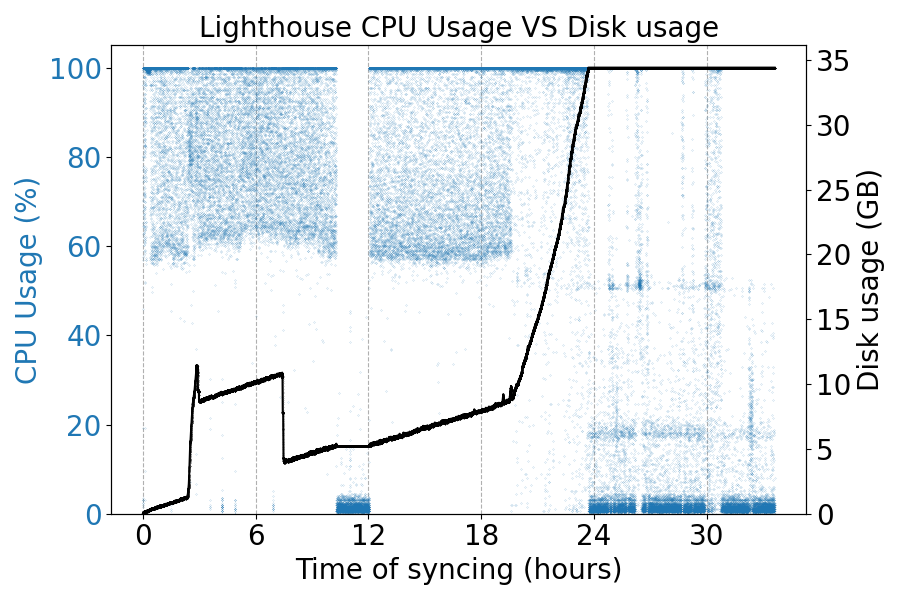}}
\caption{Lighthouse CPU (blue) and disk usage (black)}
\label{fig:lighthouse-cpu-disk}
\end{figure}

In figure~\ref{fig:lighthouse-cpu-disk}, we compare the disk and CPU utilization of Lighthouse. As we can see, in the same spots where there is no CPU utilization (i.e., hours 10-12 and 24-34) the disk utilization also becomes flat, showing that the client stall during those last hours.

This phenomena, could be linked to a Medalla non-finality period during October-November caused by a low number of validators\cite{medalla-nonfinality-october}. With the exception of Lighthouse's behaviour, possibly caused by a consensus problem in the Medalla Testnet, Lodestar seems to use more storage than all the other clients. In addition, it can be noticed that Lodestar starts gathering data after a long period of time (4 hours), which seems to be linked to the time spent getting the genesis block.

Teku, on the other hand, demonstrated a strange disk usage pattern that resembles to a zig-zag as shown in Figure~\ref{fig:clients-disk}. This zig-zag pattern could be explained by the fact that Teku uses RocksDB, which sometimes implements a periodic compression scheme in which the database grows for a while and then compacts itself. Nonetheless, its periodicity gets disrupted at some point, and from hours 4 to 6 we can see an important perturbation in the disk utilization.

\begin{figure}[h]
\centerline{\includegraphics[scale=0.38]{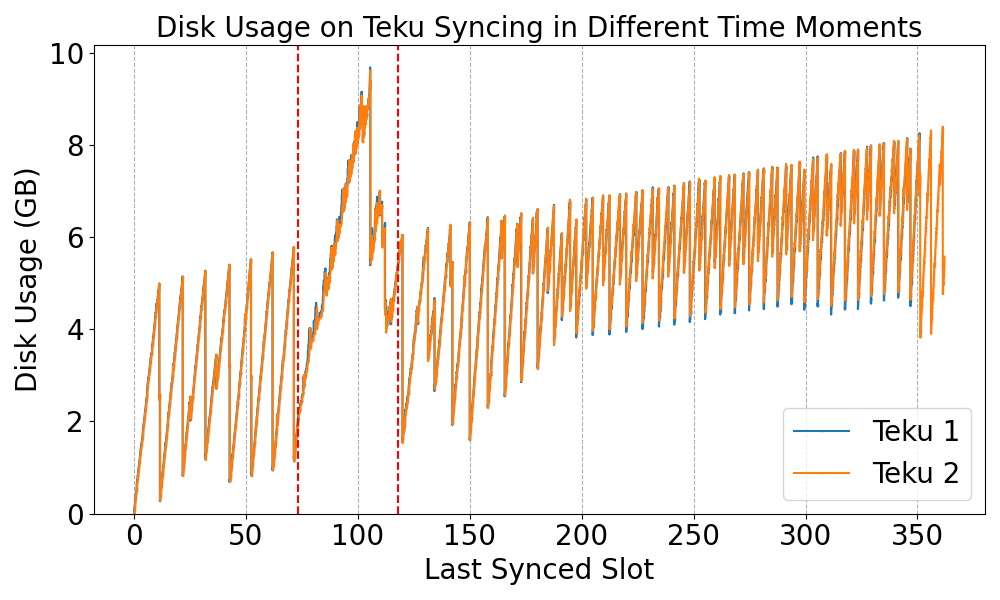}}
\caption{Teku Disk Usage for two different executions}
\label{fig:tekuDISK-Current_SlotComparison}
\end{figure}

\begin{figure}[h]
\centerline{\includegraphics[scale=0.38]{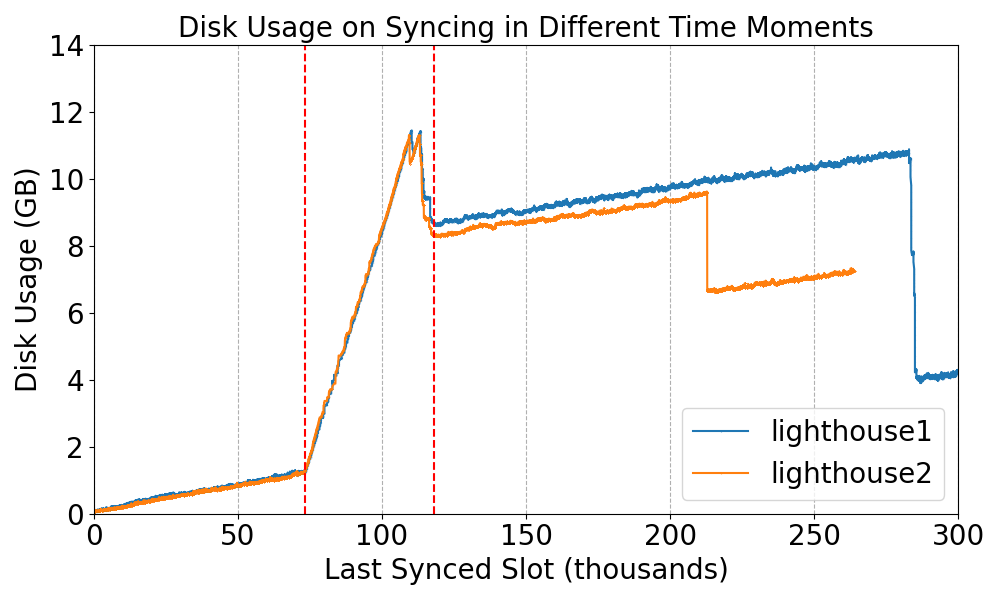}}
\caption{Lighthouse Disk Usage for two different executions}
\label{fig:lighthouseDISK-blockSlotComparison}
\end{figure}

To have a better view of the perturbation, we plotted the disk utilization of Teku by last synced slot (instead of wall-clock time) and we can clearly see that the perturbation happens between slots 70,000 and 120,000, as shown in Figure~\ref{fig:tekuDISK-Current_SlotComparison}. Given that this perturbation could be related to some random internal process in the node or events linked to RocksDB or the JVM, we decided to run Teku again from scratch on the same node. To our surprise the disk usage pattern of the second run is almost identical to the first one, with the same perturbation between the same slots (i.e., 70,000-120,000). Please note that these two runs were done one three days after the other. This deterministic behaviour seems to imply that the client is simply requesting and processing blocks in order, one after the other, which could imply that the client is syncing through a finalised part of the chain where there are no forks to explore and everything is deterministic. We also ran Lighthouse twice to check if the disk behaviour was deterministic and it is for most of the period between slots 75,000 and 120,000, as shown in Figure~\ref{fig:lighthouseDISK-blockSlotComparison}. However, Lighthouse behaviour does change a bit after slot 210,000: the second run managed to reduce storage usage way before the first run. Although, this shows that this behaviour is not just an artifact of our experiments, this does not explain why the disk grows so fast and what is the root-cause of the perturbation.

\subsection{Fossil Records of Non-Finality}
\label{sec:fossil}

\begin{figure}[h]
\centerline{\includegraphics[scale=0.38]{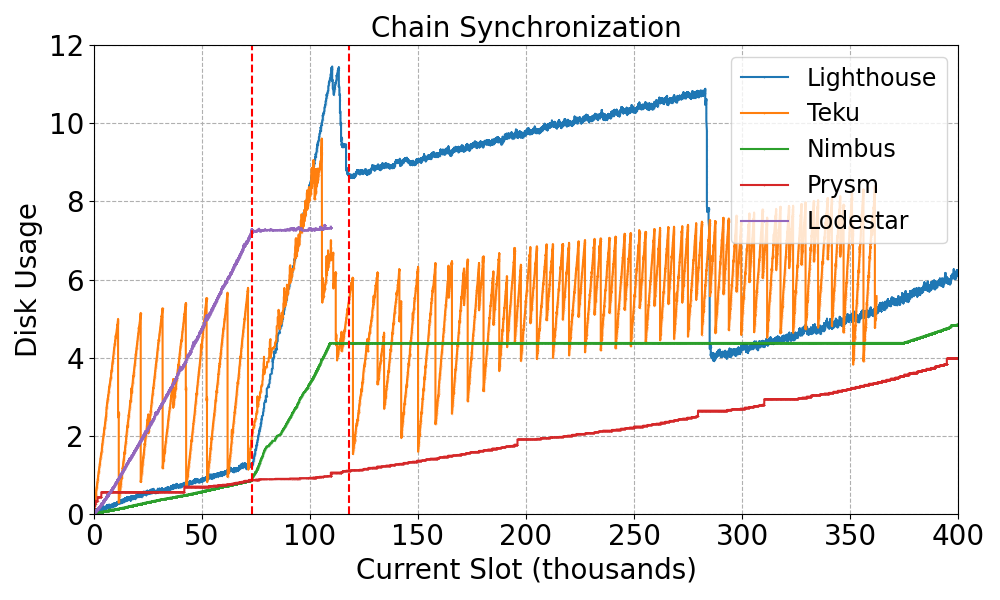}}
\caption{Disk Usage of Eth2 clients for slots}
\label{fig:client-disk-slot}
\end{figure}

In an attempt to compare the disk behaviour of all clients in a synchronous way, we have plotted the storage usage of all clients with respect to the slot number in Figure~\ref{fig:client-disk-slot}. In this plot, we see that almost all clients change behaviour somewhere around slot 70,000 and go back to a normal behaviour around slot 120,000. This period corresponds with the non-finality period of Medalla. Indeed, the Medalla network passed through a period of non-finality caused by erroneous rough time responses witnessed by Prysm clients~\cite{medalla-nonfinality-august}.

During this period, we observe a sharp increase in the disk usage of Teku, Lighthouse and Nimbus, while Lodestar shows the exact opposite and the behaviour of Prysm seems to be unaffected. For Lighthouse and Teku, we can see that the sharp increase in Disk usage is followed by a similarly sharp drop towards the end of this period. Considering the relatively similar behaviour of both clients, it is interesting to notice how Teku reduced the time of higher disk usage, while Lighthouse keeps running several hours with additional data before dumping it. This is due to the dual-database system that Lighthouse and some other clients use: a \emph{Hot} database that stores unfinalized BeaconState objects, and a \emph{cold} database that stores finalized BeaconStates. As they sync through a large patch of non-finality, their hot databases grow large until they reach finality and then migrate this state into the cold database.

\begin{figure}[H]
\centerline{\includegraphics[scale=0.36]{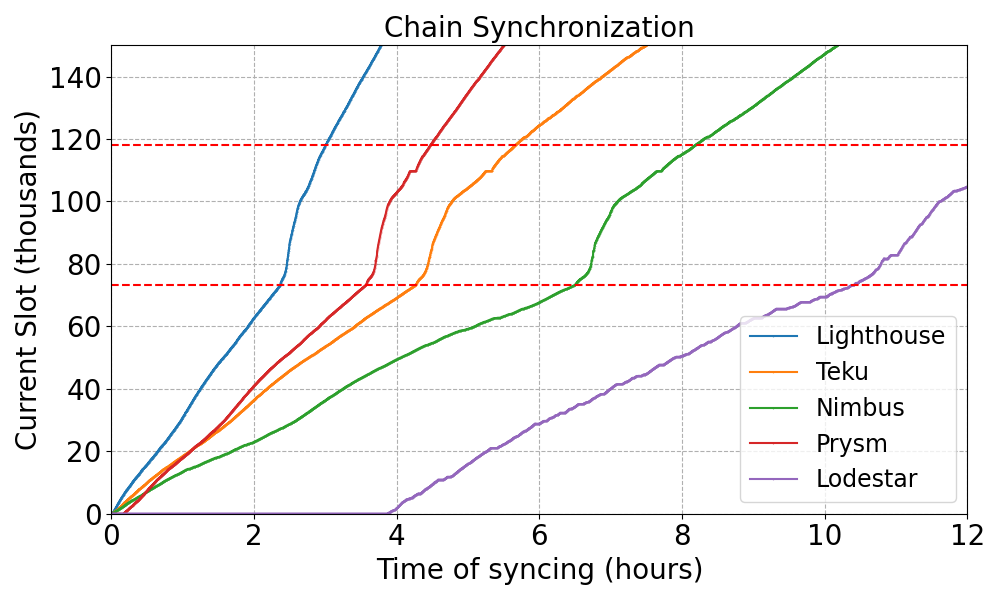}}
\caption{Detail of slot synchronization of Eth2 clients}
\label{fig:clients-current-slot-ranged}
\end{figure}

On the other hand, Nimbus rise in disk storage is not as sharp as Teku and Lighthouse, however it did not reduce its storage afterwards (in contrast to Teku and Lighthouse). Oddly, we can notice that Lodestar's disk usage increases more rapidly than any other client at the beginning, until the start of this non-finality period, when it stops growing at all. Prysm's disk usage continues its trend without any variations as if it was not perturbed by the non-finality period. This is because Prysm clients only save finalized states in intervals of every 2048 slots. This keeps disk utilization to a minimum. During non-finality, they do not save unfinalized states to disk which allows them to prevent the database from growing unnecessarily large.  However doing this comes at a cost, as they now keep everything in memory so if they do need to retrieve a particular state (unfinalized) and it's been a while since finality, they have to regenerate it. Doing this puts a non trivial amount of pressure on the CPU and can make keeping track of all the different forks harder.

\begin{figure}[H]
\centerline{\includegraphics[scale=0.36]{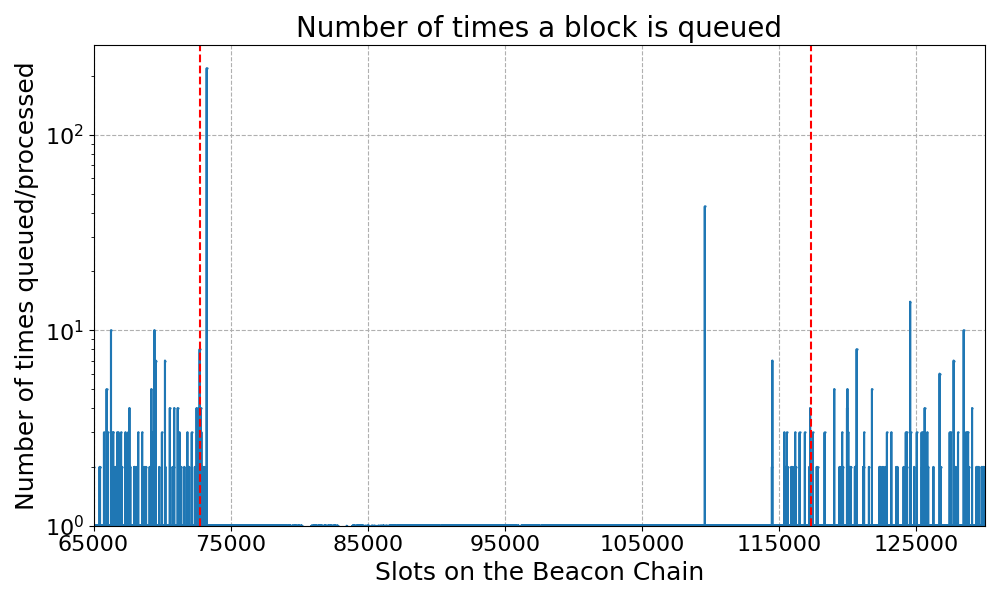}}
\caption{Number of times a block is queued}
\label{fig:clients-slot}
\end{figure}

During this non-finality period, clients also showed a steeper syncing curve around slot 100,000, as we can see in Figure \ref{fig:clients-current-slot-ranged}. This could imply that during this period there is little information to process and therefore clients can move faster in the syncing process. However, this does not seem to fit with the accelerating disk usage observed during the same period. To look deeper into this question, we used Teku logs to analyse the number of times a block was queued and/or processed for each slot during the non-finality period. The results, depicted in Figure~\ref{fig:clients-slot}, show that during this period there were almost no blocks queued, which seems to be consistent with the accelerated syncing speed. However we also notice that just at the beginning of the non-finality period, at exactly slot 73,248, there were 219 blocks queued (note the logarithmic Y axis). This clearly shows a huge perturbation in the network.

\begin{figure}[H]
\centerline{\includegraphics[scale=0.32]{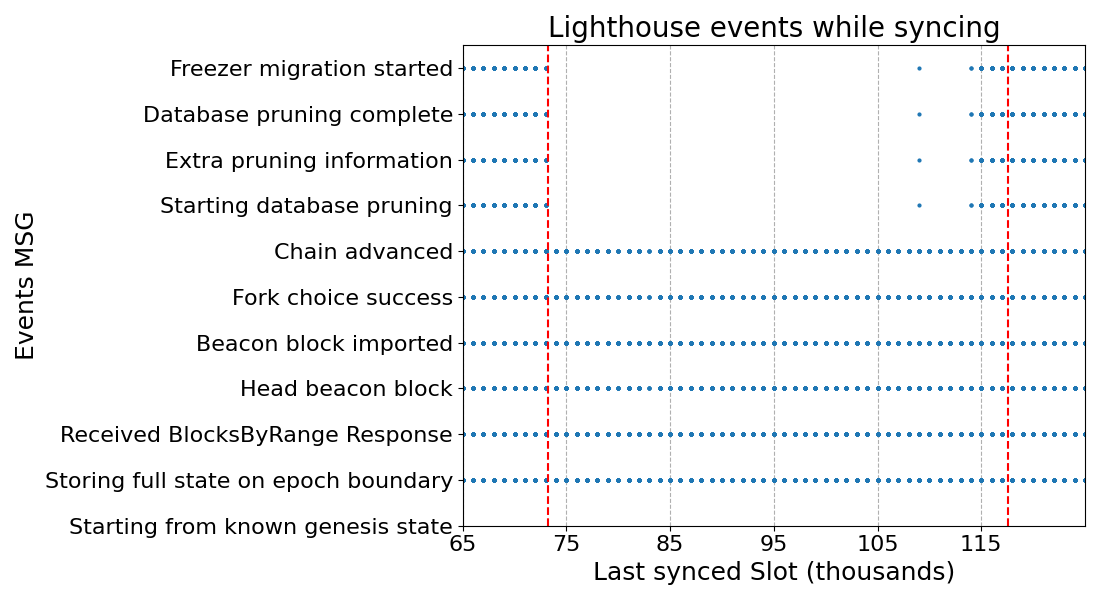}}
\caption{Events timeline for Lighthouse}
\label{fig:clients-current-slot-zoomed}
\end{figure}

We assume that the accelerating disk usage is related to an increase in the state stored in the database of the client, and this might be linked to a difficulty of pruning states during a non-finality period. Thus, to corroborate our hypothesis, we analysed Lighthouse detailed logs and plot the frequency at which different events get executed. Figure~\ref{fig:clients-current-slot-zoomed} lists a total of 11 different types of events. We can see that during the non-finality period there are four type of events that almost never get executed: \emph{Freezer migration started, Database pruning complete, Extra pruning information, and Starting database pruning}. This demonstrates that during this period the client is unable to prune the database, which is consistent with the rapid disk usage increase.

Although there are multiple of things that remain to be understood about the behaviour of some clients during a non-finality period, we have demonstrated in this paper that it is possible to identify such a network disturbance by simply looking at the resource utilization of the clients.

\section{Related Work}
\label{sec:relatedwork}

In recent years, there have been significant efforts to study the scalability and security of the Ethereum network from multiple perspectives. The importance of a new consensus mechanism for Ethereum can be understood through the scalability challenges that the protocol faces~\cite{8705874}. Peers participation of Ethereum network has been studied previously~\cite{10.1145/3278532.3278542}.

Eth2 clients rely on the Proof-of-Stake (PoS) consensus mechanism called Casper~\cite{buterin2019casper} and some works have shown insights on the Casper's robustness depending on the network latency \cite{Moindrot2017ProofOS}. However, Casper and the Beacon Chain is just the first step (i.e., Phase 0) of a fully sharded Eth2 protocol, and its following phase concerns the problem of data availability~\cite{10.1145/3284764.3284769}. In addition, the current security threats of Ethereum have also been analysed recently~\cite{10.1145/3391195}.
While all these analyses are very valuable for the community, they all focus on the protocol security aspects and none of them evaluates the actual implementation of the protocol.

Eth2 consensus mechanism requires validators to lock ETH into a deposit smart-contract, whose security is essential. In order to verify and increase the security of the deposit contract, formal verification of the Eth2 deposit contract has been done~\cite{10.1007/978-3-030-53288-8_8}. This study does verify the implementation of the deposit contract but not the Eth2 clients. While Eth1 clients have also been analysed in the past~\cite{8342866}, to the best of our knowledge there is no a study showing the resource utilization of Eth2 clients and what can be understood from it.

%As understanding the performance of several Eth2 clients is relevant to the security of the network, it is useful to get 

%It is also interesting to give a context to this new generation of clients that we have analysed in this study, by can be referencing the analysis made on the 

%Furthermore, as the study that we have made analysed the performances and behaviour of Eth2 clients when interacting with the Beacon Chain, it is interesting to remember that 

\section{Conclusion}
\label{sec:conclusion}

In this work, we have performed a detailed analysis of the resource usage of Eth2 clients. Through this analysis, we have been able to notice the effort of different teams to develop functional software in order to tackle the transitioning to Eth2. The necessity of relying on several types of clients is reflected by the different objectives and resources of a broad variety of users. Our analysis showed significant differences between clients in terms of CPU, memory and disk utilization. We also observed some network perturbations where the number of peers dropped quickly and its impact on the other resources. 
To the best of our knowledge, this is the first study of the Eth2 clients resource usage at this scale and to demonstrate that it is possible to detect traces of non-finality periods by simply looking at the clients' resource usage. The diverse behaviour of the analysed clients is a positive sign for the Ethereum community, as it shows a solid variety of software available for Eth2 validators, and this provides the network with further decentralization, resilience and security.

As future work, we would like to further study how non-finality periods affect the resources usage of Eth2 clients, in order to implement online detectors that use this information to signal possible network perturbations in real time.

\section*{Acknowledgment}

This work has been supported by the Ethereum Foundation under Grant FY20-0198. We would like to thank the researchers of the Ethereum Foundation for their feedback and suggestions. We would also like to thank the developers and researchers of the five client teams for all their help to set up our experiments and their constructive feedback on this study.  

\bibliographystyle{IEEEtran}
\bibliography{icbc}

\begin{thebibliography}{10}
\providecommand{\url}[1]{#1}
\csname url@samestyle\endcsname
\providecommand{\newblock}{\relax}
\providecommand{\bibinfo}[2]{#2}
\providecommand{\BIBentrySTDinterwordspacing}{\spaceskip=0pt\relax}
\providecommand{\BIBentryALTinterwordstretchfactor}{4}
\providecommand{\BIBentryALTinterwordspacing}{\spaceskip=\fontdimen2\font plus
\BIBentryALTinterwordstretchfactor\fontdimen3\font minus
  \fontdimen4\font\relax}
\providecommand{\BIBforeignlanguage}[2]{{%
\expandafter\ifx\csname l@#1\endcsname\relax
\typeout{** WARNING: IEEEtran.bst: No hyphenation pattern has been}%
\typeout{** loaded for the language `#1'. Using the pattern for}%
\typeout{** the default language instead.}%
\else
\language=\csname l@#1\endcsname
\fi
#2}}
\providecommand{\BIBdecl}{\relax}
\BIBdecl

\bibitem{eth-whitepaper}
``Ethereum whitepaper,'' \url{ https://ethereum.org/en/whitepaper/}.

\bibitem{phase-0}
``Phase-0,'' \url{ https://notes.ethereum.org/@djrtwo/Bkn3zpwxB}.

\bibitem{medallanet}
``Medalla test network:,'' \url{ https://github.com/goerli/medalla}.

\bibitem{metrics-monitoring-scripts}
``Monitoring scripts:,''
  \url{https://github.com/leobago/BSC-ETH2/tree/master/ETH2-clients-comparison/scripts/parsing_scripts/metrics_monitoring_scripts}.

\bibitem{logs-parsing-scripts}
``Logs parsing scripts:,''
  \url{https://github.com/leobago/BSC-ETH2/tree/master/ETH2-clients-comparison/scripts/parsing_scripts/client_logs_analysis_scripts}.

\bibitem{plotting-scripts}
``Plotting scripts:,''
  \url{https://github.com/leobago/BSC-ETH2/tree/master/ETH2-clients-comparison/scripts/plotting_scripts}.

\bibitem{teku-version}
``Teku version:,'' \url{
  https://github.com/ConsenSys/teku/commit/6883451c9f87a28c5923dbc8f291db237930cad0}.

\bibitem{prysm-version}
``Prysm version:,'' \url{
  https://github.com/prysmaticlabs/prysm/commit/4bc7cb6959a1ea5b4b4b53b42284900e3b117dea}.

\bibitem{lighthouse-version}
``Lighthouse version:,'' \url{
  https://github.com/sigp/lighthouse/commit/95c96ac567474df2abb4e9da9f5e771cf5a7426d}.

\bibitem{nimbus-version}
``Nimbus version:,'' \url{
  https://github.com/status-im/nimbus-eth2/commit/9255945fb0ef1bf036d32b7cce5df42a8dd69be7}.

\bibitem{lodestar-version}
``Lodestar version:,'' \url{ https://github.com/ChainSafe/lodestar/pull/1731}.

\bibitem{afri}
``Multi-client benchmark on medalla testnet 2020/10/01:,''
  \url{https://github.com/q9f/eth2-bench-2020-10}.

\bibitem{medalla-nonfinality-october}
``Medalla non-finality october 2020:,''
  \url{https://gist.github.com/yorickdowne/ea9b18ac2b51a508080c9a810d978522}.

\bibitem{medalla-nonfinality-august}
``Medalla non-finality period august 2020:,''
  \url{https://docs.google.com/document/d/11RmitNRui10LcLCyoXY6B1INCZZKq30gEU6BEg3EWfk}.

\bibitem{8705874}
M.~{Bez}, G.~{Fornari}, and T.~{Vardanega}, ``The scalability challenge of
  ethereum: An initial quantitative analysis,'' in \emph{2019 IEEE
  International Conference on Service-Oriented System Engineering (SOSE)},
  April 2019, pp. 167--176.

\bibitem{10.1145/3278532.3278542}
\BIBentryALTinterwordspacing
S.~K. Kim, Z.~Ma, S.~Murali, J.~Mason, A.~Miller, and M.~Bailey, ``Measuring
  ethereum network peers,'' in \emph{Proceedings of the Internet Measurement
  Conference 2018}, ser. IMC '18.\hskip 1em plus 0.5em minus 0.4em\relax New
  York, NY, USA: Association for Computing Machinery, 2018, p. 91–104.
  [Online]. Available: \url{https://doi.org/10.1145/3278532.3278542}
\BIBentrySTDinterwordspacing

\bibitem{buterin2019casper}
V.~Buterin and V.~Griffith, ``Casper the friendly finality gadget,'' 2019.

\bibitem{Moindrot2017ProofOS}
O.~Moindrot, ``Proof of stake made simple with casper,'' 2017.

\bibitem{10.1145/3284764.3284769}
\BIBentryALTinterwordspacing
D.~Sel, K.~Zhang, and H.-A. Jacobsen, ``Towards solving the data availability
  problem for sharded ethereum,'' in \emph{Proceedings of the 2nd Workshop on
  Scalable and Resilient Infrastructures for Distributed Ledgers}, ser.
  SERIAL'18.\hskip 1em plus 0.5em minus 0.4em\relax New York, NY, USA:
  Association for Computing Machinery, 2018, p. 25–30. [Online]. Available:
  \url{https://doi.org/10.1145/3284764.3284769}
\BIBentrySTDinterwordspacing

\bibitem{10.1145/3391195}
\BIBentryALTinterwordspacing
H.~Chen, M.~Pendleton, L.~Njilla, and S.~Xu, ``A survey on ethereum systems
  security: Vulnerabilities, attacks, and defenses,'' \emph{ACM Comput. Surv.},
  vol.~53, no.~3, Jun. 2020. [Online]. Available:
  \url{https://doi.org/10.1145/3391195}
\BIBentrySTDinterwordspacing

\bibitem{10.1007/978-3-030-53288-8_8}
D.~Park, Y.~Zhang, and G.~Rosu, ``End-to-end formal verification of ethereum
  2.0 deposit smart contract,'' in \emph{Computer Aided Verification}, S.~K.
  Lahiri and C.~Wang, Eds.\hskip 1em plus 0.5em minus 0.4em\relax Cham:
  Springer International Publishing, 2020, pp. 151--164.

\bibitem{8342866}
S.~{Rouhani} and R.~{Deters}, ``Performance analysis of ethereum transactions
  in private blockchain,'' in \emph{2017 8th IEEE International Conference on
  Software Engineering and Service Science (ICSESS)}, 2017, pp. 70--74.

\end{thebibliography}

\end{document}